%% file: main.tex
\documentclass[conference]{IEEEtran}
\IEEEoverridecommandlockouts

\input{packages}

\begin{document}

\input{title}

\input{introduction}

\input{channel_sounder_setup}

\input{measurement_scenario}

\input{evaluation}



\input{bibliography}
\end{document}

%% file: packages.tex
\usepackage{cite}
\usepackage{amsmath,amssymb,amsfonts}
\usepackage{algorithmic}
\usepackage{graphicx}
\usepackage{textcomp}
\usepackage{xcolor}
\usepackage{comment}
\usepackage{float}
\def\BibTeX{{\rm B\kern-.05em{\sc i\kern-.025em b}\kern-.08em
    T\kern-.1667em\lower.7ex\hbox{E}\kern-.125emX}}
    
\usepackage{siunitx}
    \DeclareSIUnit{\belmilliwatt}{Bm}
    \DeclareSIUnit{\belisotropic}{Bi}
    \DeclareSIUnit{\rpm}{rpm}

%% file: title.tex
\title{Sub-THz Channel Measurements at 158 GHz and 300 GHz in a Street Canyon Environment}

\author{\IEEEauthorblockN{
Wilhelm Keusgen\IEEEauthorrefmark{1}, 
Alper Schultze\IEEEauthorrefmark{2}, 
Michael Peter\IEEEauthorrefmark{2}, 
Taro Eichler\IEEEauthorrefmark{3}}

\IEEEauthorblockA{
\IEEEauthorrefmark{1}
Technische Universität Berlin, Berlin, Germany, \emph{wilhelm.keusgen@tu-berlin.de}}
\IEEEauthorblockA{
\IEEEauthorrefmark{2}
Fraunhofer Heinrich Hertz Institute, Berlin, Germany}
\IEEEauthorblockA{
\IEEEauthorrefmark{3}
Rohde \& Schwarz, Munich, Germany}}

\maketitle

%% file: introduction.tex
\section{Introduction}
This paper presents first results of a channel measurement campaign performed in an urban micro (UMi) street canyon scenario at \SI{158}{\giga\hertz} and \SI{300}{\giga\hertz}. The measurements are part of a larger research activity aiming for a better understanding of the millimeter and sub-millimeter (sub-THz and THz) mobile radio channel in extension to prior work \cite{bib3}. The frequencies were chosen with respect to ongoing discussions for the sixth generation of mobile networks (6G). The presented results address fundamental questions with respect to the processing of measurement data and give some insight into typical properties of the radio channel at these frequencies.

%% file: channel_sounder_setup.tex
\section{Channel Sounder Setup}

The measurements were captured using an instrument-based time-domain channel sounder, equipped with D-Band (\SI{110}{\giga\hertz} to \SI{170}{\giga\hertz}) and H-Band (\SI{220}{\giga\hertz} to \SI{330}{\giga\hertz}) front-ends. The setup consists of a static transmitter and a mobile receiver. The transmitter comprises a broad-band vector signal generator, that generates an IF signal from a digital baseband sequence, as well as a single-sideband upconverter with a distinct LO source, that mixes the signal into the RF domain. The receiver consists of a step-wise rotatable downconverter with a distinct LO source that amplifies the antenna signal and mixes it into an IF domain, and a signal analyzer that samples the IF signal and stores it as IQ samples. At each measurement point, the receiver was rotated by 360 degrees in 24 uniform steps, resulting in a directional sampling of the propagation channel in the azimuth plane with a resolution of 15 degrees. For each angle approached, 150 sequence snapshots were taken and used for averaging to increase the instantaneous dynamic range. For both bands, the receiver was equipped with  standard gain horn antennas with a gain of \SI{20}{\deci\belisotropic}. Transmitter and receiver are synchronized with two rubidium-based reference clocks and trigger units, enabling coherent measurements and the determination of the absolute time of flight. The fundamental channel sounder parameters are the carrier frequencies of \SI{158}{\giga\hertz} and \SI{300}{\giga\hertz}, a measurement bandwidth of \SI{2}{\giga\hertz}, and the use of a complex correlation sequence with a duration of \SI{100}{\micro\second}. Further information about the setup and performed measurements at \SI{300}{\giga\hertz} can be found in \cite{bib1, bib2}.

%% file: measurement_scenario.tex
\section{Measurement Scenario and Procedure}

\begin{figure}[htbp]
\includegraphics[width=\columnwidth]{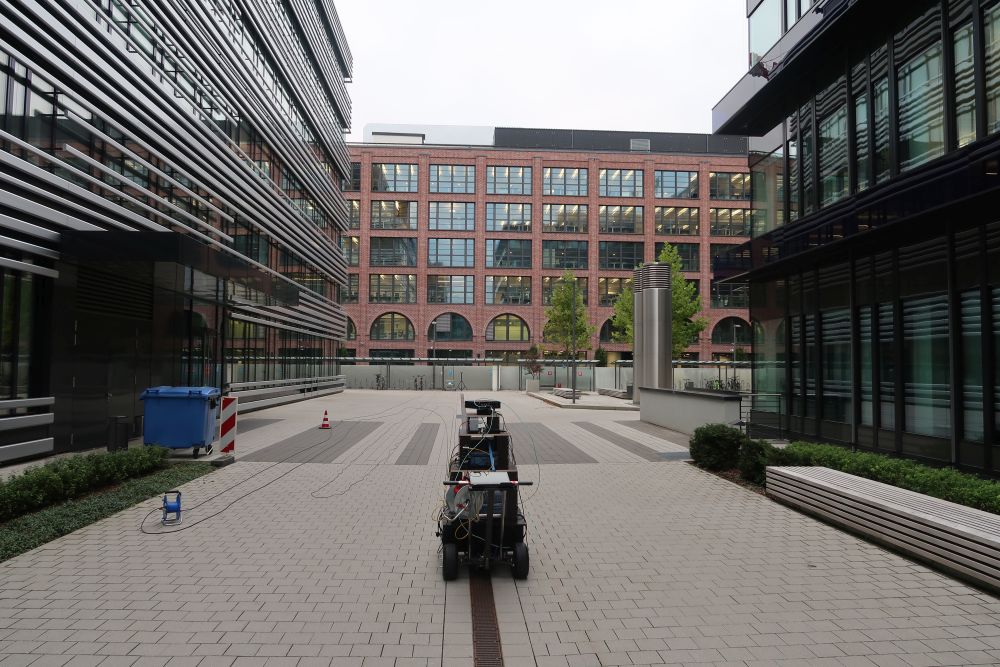}
\caption{UMi street canyon measurement scenario showing the receiver setup in front.}
\label{fig:photo}
\end{figure}

The measurements took place on company premises in Munich, Germany, well representing a UMi street canyon scenario as shown in Fig. \ref{fig:photo}. The \SI{15.5}{\meter} wide street canyon is bordered by two buildings with a height of approx. \SI{20}{\meter}. 

The transmitter was located in the middle of the street at a height of \SI{1.5}{\meter}. The receiver, which exhibited the same antenna height, was moved to multiple positions along the middle of the street up to a maximum distance of \SI{170}{\meter}. Most of the measurements were made in line-of-sight condition.

%% file: evaluation.tex
\section{Measurement Evaluation and Results}

\begin{figure*}[htbp]
    \centering
    \begin{minipage}{.48\textwidth}
        \includegraphics[scale=1]{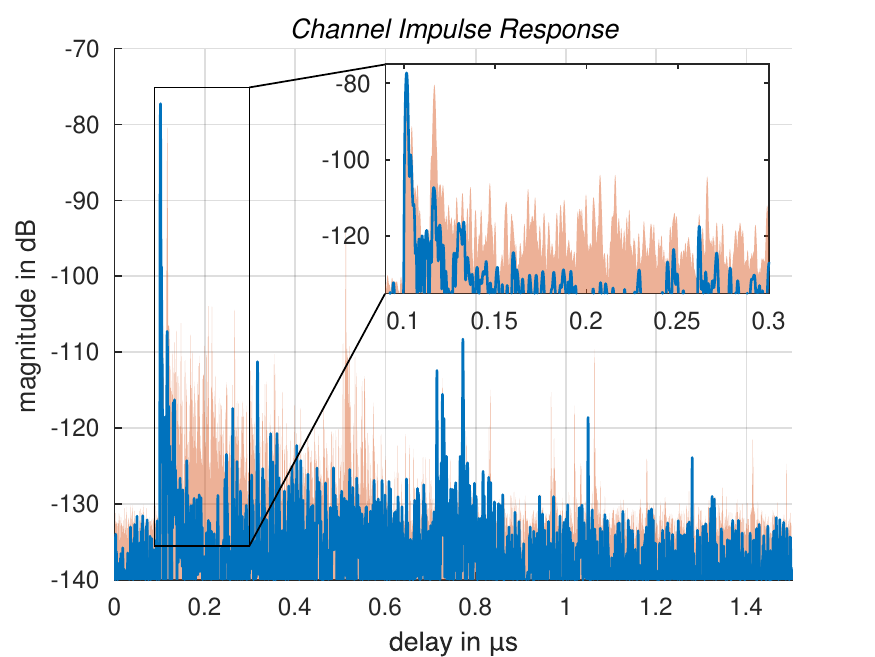}
    \end{minipage} \quad
    \begin{minipage}{.48\textwidth}
        \includegraphics[scale=1]{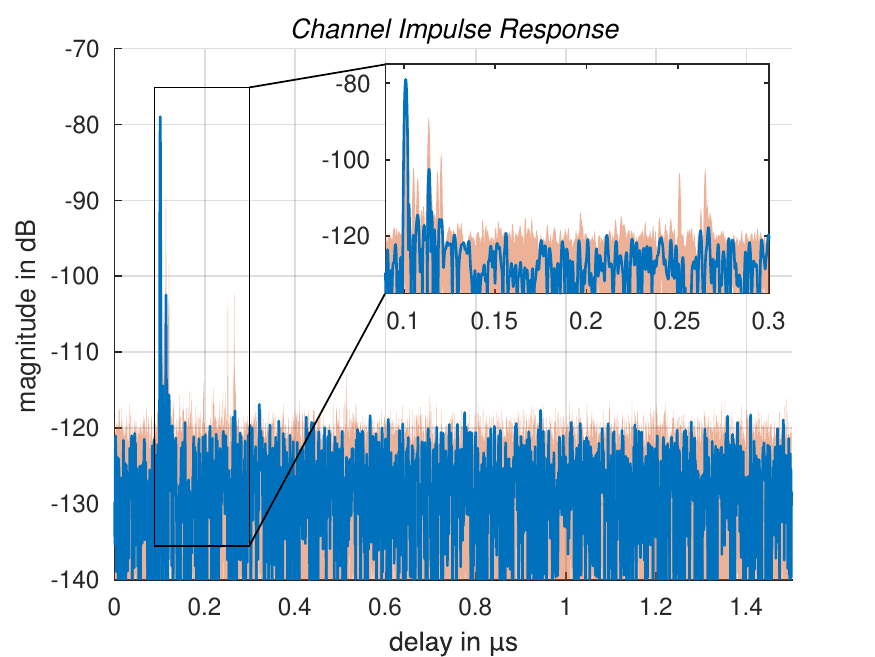}
    \end{minipage} \quad
    \caption{CIRs in the direction of the LOS path and pseudo-omnidirectional CIRs for \SI{158}{\giga\hertz} (left) and \SI{300}{\giga\hertz} (right).}
    \label{fig:cir}
\end{figure*}
\begin{figure*}[htbp]
    \vspace{-0.3cm}
    \begin{minipage}{.47\textwidth}
        \includegraphics[scale=1]{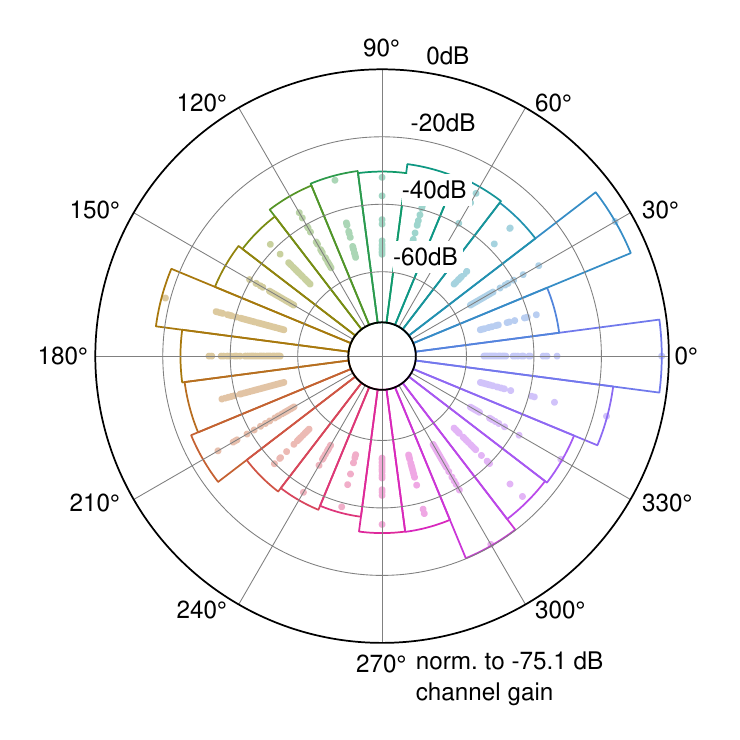}
    \end{minipage} \quad
    \begin{minipage}{.47\textwidth}
        \includegraphics[scale=1]{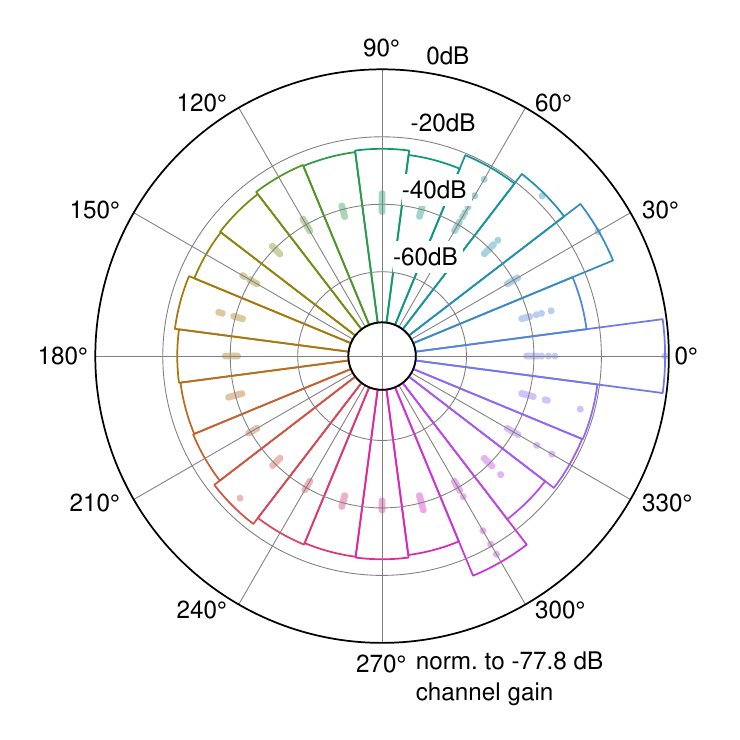}
    \end{minipage} \quad
    \vspace{-0.5cm}
    \caption{Rose plots for \SI{158}{\giga\hertz} (left) and \SI{300}{\giga\hertz} (right).}
    \label{fig:rose}
\end{figure*}

In order to get calibrated channel impulse responses (CIRs), the accumulated IQ samples are passed through several processing steps, which include resampling, filtering, estimation and compensation of common phase drift, coherent averaging and correlation with pre-recorded back-to-back calibration data. 
Fig. \ref{fig:cir} shows measurement results for both carrier frequencies at a distance of \SI{30}{\meter}. Both figures illustrate the CIR measured in the direction of the LOS path as a line plot and the pseudo-omnidirectional CIR including all directions as an area plot. The leading part of the CIR is additionally depicted with an expanded delay axis.

One can see the first peaks at \SI{0.1}{\micro\second}, which corresponds to a distance of \SI{30}{\meter}. The noise floor at \SI{158}{\giga\hertz} can be estimated to around \SI{-140}{\deci\bel} (mean normalized power), whereas the noise floor for \SI{300}{\giga\hertz} is around \SI{-130}{\deci\bel}. This difference corresponds to the unequal transmit powers (\SI{10}{\deci\belmilliwatt} versus \SI{3}{\deci\belmilliwatt}) and the higher noise figure of the receiver at 300 GHz. The channel gains of the strongest path are \SI{-77.3}{\deci\bel} and \SI{-79.0}{\deci\bel} (\SI{158}{\giga\hertz} and \SI{300}{\giga\hertz}), whereas the theoretical values from free-space path loss and including antenna gains are \SI{-78.0}{\deci\bel} and \SI{-83.5}{\deci\bel}. The slightly lower path loss at \SI{300}{\giga\hertz} may be caused by a ground reflection that coincidentally superimposes constructively on the direct path and needs further investigation.

Further evaluations of the measured data sets comprise the investigation of the direction-of-arrival information. For this, an estimation of distinct propagation paths is performed by means of a local peak search in the angular-delay domain \cite{bib2}, whereas the angular dimension is spanned by the $24$ different measurement directions (angle bins). Fig. \ref{fig:rose} shows an representation of the results in the form of rose plots. Here each “wedge” represents the power in the respective angle bin, normalized to the total channel gain i.e. the sum power of all estimated paths. The individual contributions from distinct paths within each angle bin are depicted by dots. It can clearly be seen that only one or two angle bins account for almost all of the total power, and that within each bin only a few paths make a significant contribution.

%% file: main.bbl
\begin{thebibliography}{unsrt}

\bibitem{bib3}J. M. Eckhardt, T. Doeker, S. Rey and T. Kürner, "Measurements in a Real Data Centre at 300 GHz and Recent Results," 2019 13th European Conference on Antennas and Propagation (EuCAP), 2019, pp. 1-5.
\bibitem{bib1} M. Schmieder et al., "THz Channel Sounding: Design and Validation of a High Performance Channel Sounder at 300 GHz," 2020 IEEE Wireless Communications and Networking Conference Workshops (WCNCW), 2020, pp. 1-6, doi: 10.1109/WCNCW48565.2020.9124887.
\bibitem{bib2} F. Undi, A. Schultze, W. Keusgen, M. Peter, and T. Eichler, "Angle-Resolved THz Channel Measurements at 300 GHz in an Outdoor Environment," 2021 IEEE International Conference on Communications Workshops (ICC Workshops), 2021, pp. 1-7, doi: 10.1109/ICCWorkshops50388.2021.9473891.


\end{thebibliography}
